\documentclass[conference,letterpaper]{IEEEtran}
\IEEEoverridecommandlockouts
\usepackage{cite}
\usepackage{amsmath,amssymb,amsfonts}
\usepackage{algorithmic}
\usepackage{graphicx}
\usepackage{textcomp}

\usepackage{CJKutf8}
\begin{CJK*}{UTF8}{gbsn}

\usepackage{setspace}

\def\BibTeX{{\rm B\kern-.05em{\sc i\kern-.025em b}\kern-.08em
    T\kern-.1667em\lower.7ex\hbox{E}\kern-.125emX}}

\begin{document}

\title{The Coming Era of AlphaHacking?\protect\\[0.8ex]
\begin{LARGE}
\begin{spacing}{1.0}
A Survey of Automatic Software Vulnerability Detection,\protect\\Exploitation and Patching Techniques
\end{spacing}
\end{LARGE}
}

\author{\IEEEauthorblockN{Tiantian Ji\IEEEauthorrefmark{1},Yue Wu\IEEEauthorrefmark{1},Chang Wang\IEEEauthorrefmark{1}, Xi Zhang\IEEEauthorrefmark{1},Zhongru Wang\IEEEauthorrefmark{1}\IEEEauthorrefmark{2}}
\IEEEauthorblockA{\textit{\IEEEauthorrefmark{1}Key Laboratory of Trustworthy Distributed Computing and Service (BUPT), Ministry of Education,}\\
\textit{Beijing University of Posts and Telecommunications}
\IEEEauthorblockA{\textit{\IEEEauthorrefmark{2}Chinese Academy of Cyberspace Studies}}
Beijing, China
}
}

\maketitle

\begin{abstract}
With the success of the Cyber Grand Challenge (CGC) sponsored by DARPA, the topic of Autonomous Cyber Reasoning System (CRS) has recently attracted extensive attention from both industry and academia. Utilizing automated system to detect, exploit and patch software vulnerabilities seems so attractive because of its scalability and cost-efficiency  compared with the human expert based solution. In this paper, we give an extensive survey of former representative works related to the underlying technologies of a CRS, including vulnerability detection, exploitation and patching. As an important supplement, we then review several pioneer studies that explore the potential of machine learning technologies in this field, and point out that the future development of Autonomous CRS is inseparable from machine learning.
\end{abstract}


\section{Introduction}
With the rapid development of information technology, software is playing an important role in various aspects all over the world, such as the economy, military, society, etc.
At the same time,
software security is becoming an emerging worldwide challenge.
Software vulnerabilities are one of the root causes of security problems.
High skilled hackers can exploit the software vulnerabilities to do a lot of harmful things, such as stealing the private information of users, halting the crucial equipments and so on.
According to the statistics released by the Common Vulnerabilities and Exposures (CVE) organization, the number of software vulnerabilities discovered in 1999 was less than 1600 while the number of vulnerabilities currently covered by CVE and National Vulnerability Database (NVD) almost nearly 100000\cite{CVE2018}.
Facing massive software vulnerabilities, developers have neither sufficient time nor enough resources to fix all of them.
So there is the question:
which bugs are exploitable and thus should be fixed first
\cite{avgerinos2014exploiting}? Given that, the primary task in software security is to find the security-critical bugs and fix them.

However, finding and patching software vulnerabilities are highly professional works. The number of developers who can handle these jobs is far from enough compared to the number of software vulnerabilities. Therefore, the need for automated, scalable, machine-speed vulnerability detection and patching is becoming urgent when there are millions of software systems launched every day today.

To this end, DARPA held the CGC in 2016, which not only showed us an impressive scene that machines are effectively hacking against each other, but also dramatically promoted the progress of automatic software vulnerabilities detection, exploitation and patching techniques. Meanwhile, with the tremendous development of artificial intelligence announced by Google \cite{silver2016mastering}, we can see new opportunities that the software security problems can be solved by computing systems automatically and intelligently.

From this point, we give an extensive study of  former representative works related to the underlying techniques of a CRS including vulnerability detection, exploitation and patching. Although some researchers like Brooks \cite{brooks2017survey} has proposed a survey taken based on two CRS instances which are Mayhem and Mechanical Phish in this area, our work is different from theirs in several ways. First, we review these works from the perspective of both timeline and categorization. Second, as an important supplement, we then review several pioneer studies that explore the potential of machine learning technologies in this field. At last, our survey focuses on the state-of-the-art works. TABLE I gives an overview of these techniques.
\section{Automatic Vulnerability Detection}
The prevalence of vulnerabilities and the seriousness of their consequences have prompted researchers to focus more attention on digging vulnerabilities. From the perspective of whether to run the program being tested, vulnerability detection techniques can be roughly divided into three categories: static analysis, dynamic analysis and mixed analysis.

\subsection{Static Analysis}
Static analysis reasons about a program without executing it\cite{brooks2017survey}.
The general process of binary static analysis is shown in Fig. 1.
And as shown in Fig. 1, from the perspective of program modeling, we split it into two categories based on the modeling of static analysis:
graph-based static analysis and static analysis with data modeling. We then will introduce them in detail.

\subsubsection{Graph-based static analysis}
Graph-based static analysis refers to modeling program properties as graphs such as control-flow graphs (CFG), data-flow graphs (DFG) and program-dependence graphs (PDG), etc. These techniques rely on building a model of bugs by a set of nodes in the CFG or PDG to identify bugs in a program\cite{shoshitaishvili2017building}. For example, Vine, the static analysis component of BitBlaze, which provides a set of core utilities for vulnerability detection, such as CFG, DFG and weakest precondition calculation, etc\cite{song2008bitblaze}.
Yamaguchi et al.\cite{yamaguchi2014modeling} construct a code property graph that combines CFG and PDG to implement effective vulnerability detection.
Angr exposes the CFG recovery algorithms as two analyses: CFGFast and CFGAccurate\cite{shoshitaishvili2016sok}. This core component CFG provides the system with the ability to examine all possible program paths so that static analysis can achieve high coverage and also can reduce about thirty percent to seventy percent of software security issues from the root.
\begin{table}[!htbp]
\setlength{\abovecaptionskip}{0pt}
\setlength{\belowcaptionskip}{0pt}
\caption{Summary of Recent Technologies on Vulnerability Detection, Exploitation and Patching}
\includegraphics[trim=17mm 16mm 12mm 15mm, clip=true, width=3.55in]{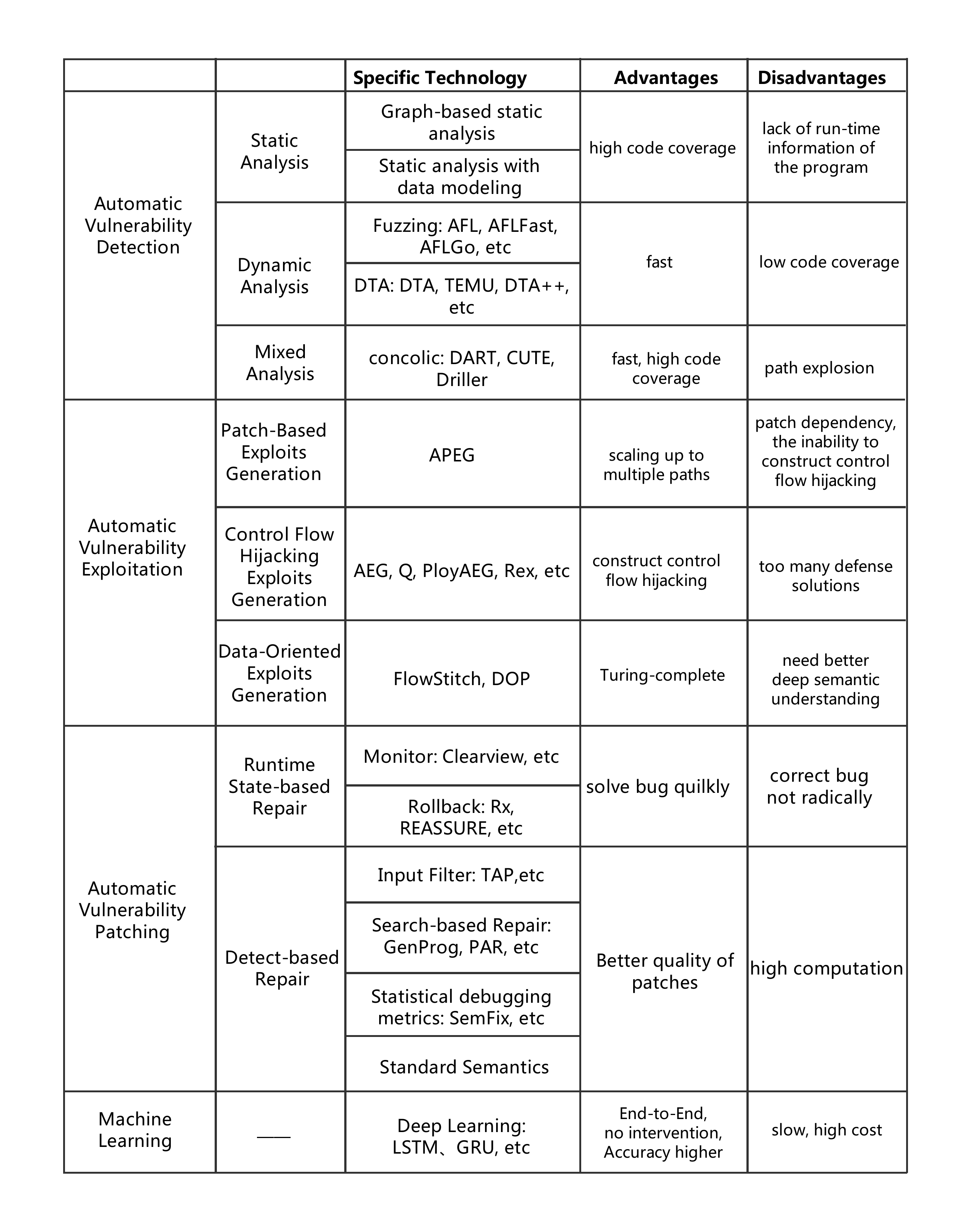}
\label{Table:1}
\end{table}

Model checking also belongs to graph-based static analysis, which refers to modeling the behavior of a program as a graph. Model checking is used in a variety of tools like slam and cboc, etc. And it has a high level of automation and complete theory but has the space-time overhead problems due to state explosion. Similarly, most methods of graph-based static analysis always meet the same problems that the solving model is too big so that the practical solution is not computable. To improve computation efficiency, researchers propose a different way of static analysis that reason over abstractions of the data upon which a program operates\cite{shoshitaishvili2017building}, which will be introduced in the following paragraph.
\begin{figure}[htbp]
\setlength{\abovecaptionskip}{0pt}
\setlength{\belowcaptionskip}{0pt}
\centering
\includegraphics[trim=8mm 15mm 8mm 15mm, clip=true, width=2.3in, height=2.36in]{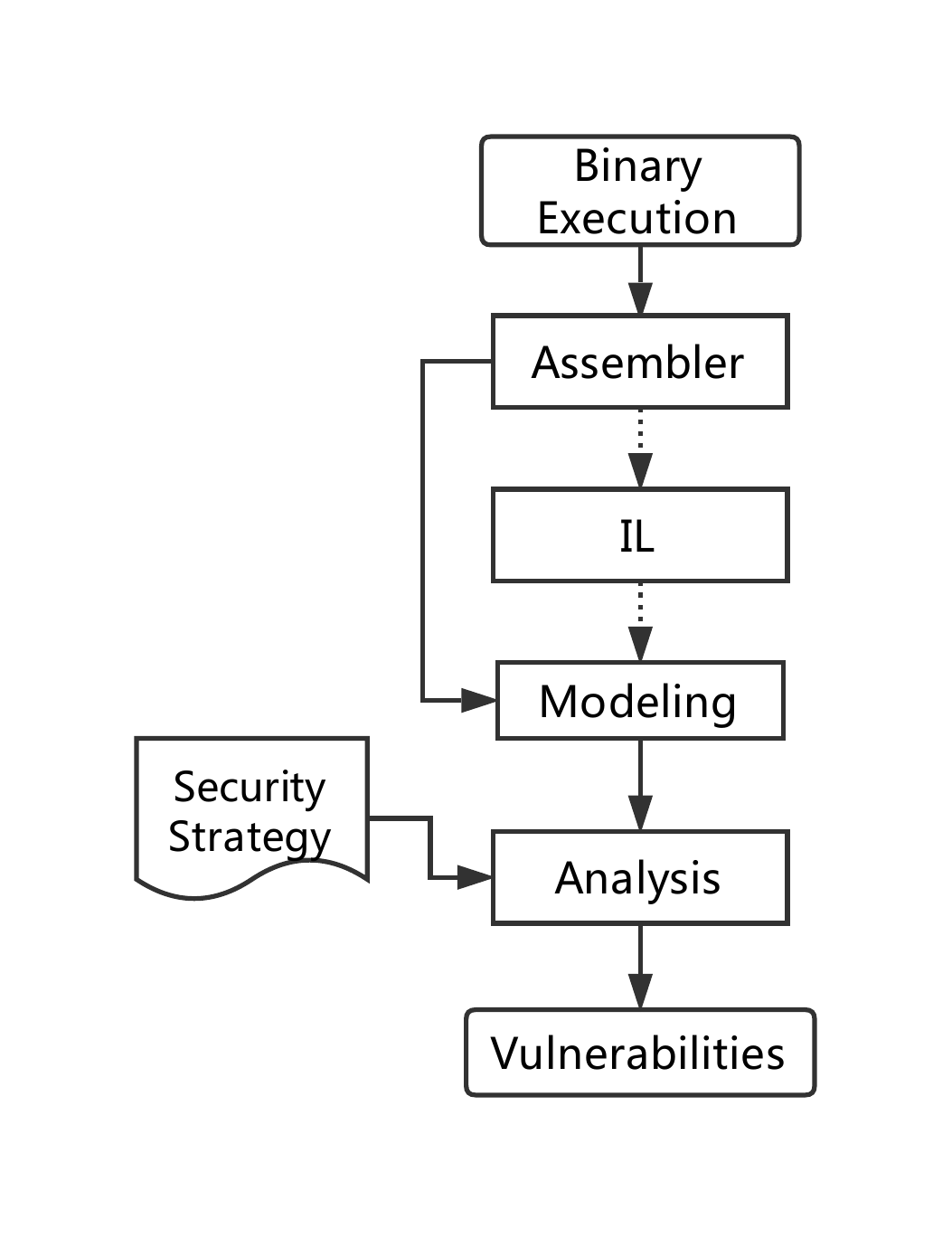}
\caption{General Process of Binary Static Analysis}
\label{Fig:1}
\end{figure}
\subsubsection{Static analysis with data modeling}
P. Cousot and R. Cousot\cite{cousot1977abstract} proposed an Abstract Interpretation (AI) method based lattice theory in 1977 to simplify and approximate the calculation of fixed points. Essentially, it is to achieve a balance between computational efficiency and computational accuracy so that AI lays a theoretical foundation for static analysis with data modeling. AI implemented in the astree static analyzer, to a certain extent, can alleviate the state explosion problem.

In this field, we consider a popular and AI-based method named Value-Set Analysis (VSA). The key to VSA is that it exceeds the approximation of the values in memory. Therefore VSA can be used to make assumptions about targets of an indirect jump statement. The original design of VSA is proposed by Balakrishnan et al.\cite{balakrishnan2008wysinwyx}. Later LoongChecker\cite{cheng2011loongchecker} uses an accurate VSA to find out addresses to solve both indirect control transfers and alias problems as many as possible. Then, Josselin Feist et al.\cite{feist2014statically} present a static tool called GUEB, which uses VSA to reason each variable in the assignment and free instructions based on an abstract memory model to search for use-after-free vulnerabilities in binary programs.

Apart from VSA, another paradigm of static analysis with data modeling is patch matching, which refers to the similarity comparison between the original program and the program with patching to get the difference part. Patch matching is a popular method in recent years. It has significant theoretical and empirical support. For example,
Li et al.\cite{li2014scalable} propose a mechanism that is syntax-based to perform well for code clone vulnerability.
Sha et al.\cite{jianming2014pvdf} present an automatic Patch-based Vulnerability Description (PVD), and it proved that all patch-related bugs can be described in PVD. Xu et al.\cite{xu2017spain} construct a patch analysis framework named SPAIN, which can automatically learn the security patch patterns and vulnerability patterns to identify and locate the vulnerability patches. Patch matching will narrow the scope of positioning, and thus it does not need to locate vulnerabilities throughout the whole program.
It also has been widely used in various tools like Klockwork, Prefix, Coverity and Fortify, etc.

The most common characteristic of the above methods is that they can achieve a high code coverage of the program, which is the advantage of static analysis. While there is also an inherent flaw in the static analysis: the lack of necessary run-time information for the program. To achieve better runtime information utilization,
dynamic analysis is put forward.

\subsection{Dynamic Analysis}
The dynamic analysis examines a program's behavior while it is running in a given environment.
Fuzzing and dynamic taint analysis are two typical dynamic analysis technologies.

\subsubsection{Fuzzing}
Fuzzing is an automated technique of black-box testing, which is a process of sending intentionally invalid data to a product in the hope of triggering an error condition or fault to determine whether there are potential security vulnerabilities\cite{sutton2007fuzzing}. Fuzzing does not require any program analysis so that it is fast and can generate multiple tests at the same time but with low coverage. As a representative of automated white-box testing, SAGE\cite{godefroid2008automated} is proposed by Microsoft Research to achieve a balance between code coverage and efficiency by repeatedly using the generational search algorithm with a coverage-maximizing heuristic design. Godefroid et al.\cite{godefroid2008automated} use it to discover more than 20 unknown vulnerabilities in large Windows applications to validate its effectiveness. But white-box fuzzing requires binary lifting and program analysis, which may result in bad efficiency. Hence researchers
propose the grey-box fuzzing which lies in between the white box analysis and the black box analysis. Without the program analysis but with more information about the internal structure, grey-box fuzzing may be more effective than white and black box analysis\cite{bohme2017coverage}.

Grey-box fuzzing is considered as the state-of-the-art technology in vulnerability detection\cite{bohme2017directed}.
American Fuzzy Lop (AFL) is its popular implementation that has detected hundreds of high-impact vulnerabilities. One of the limitations of AFL is that its mutation engine is syntax-blind, so there are many technologies trying to extend AFL. For instance, AFLFast\cite{bohme2017coverage} and AFLGo\cite{bohme2017directed} use the knowledge of Markov chain and simulated annealing algorithm respectively to extend AFL. And from the above extension of AFL, we can see a new trend that the methods based on machine learning are becoming popular and powerful in software security.

\subsubsection{Dynamic Taint Analysis}
Taint analysis was first proposed by Denning in 1976\cite{denning1976lattice}. While the Dynamic Taint Analysis (DTA) was formally proposed in 2005 by James Newsome and Dawn Song\cite{newsome2005dynamic}. DTA refers to tracking and analyzing marked information flow through the program as it executes. The reliability of DTA is verified by utilizing TaintCheck, and it is proved that DTA can detect most of the types of vulnerabilities. To improve the applicability and scalability of DTA, general frameworks are developed by researchers such as Dytan and TEMU.
However, Schwartz et al.\cite{schwartz2010all} point out that there are several fundamental challenges for DTA to achieve accurate analysis results. And the two main challenges are under-tainting and over-tainting.

Facing these problems, many optimization methods for DTA are proposed. For example, DTA++\cite{kang2011dta++} is put forward to resolve under-tainting without causing over-tainting by identifying a minimum set of implicit flows in the program that potentially cause under-tainting.
DECAF\cite{henderson2017decaf} keeps track of taint labels at bit granularity in an asynchronous manner, which enables precise and lossless dynamic taint analysis.
And also, Ma Jin-Xin et al.\cite{MaJinXin2017} pose a method based on offline indices of instruction trace to enhance DTA, and their method is verified to be efficiency. It is able to detect more vulnerabilities than TEMU, and is 5 times faster in average.

Dynamic analysis has higher accuracy than static analysis by analysing the runtime information but at the expense of less code coverage\cite{brooks2017survey}. The combination of dynamic and static analysis (i.e., mixed analysis) hereby come up with the target to achieve better analysis.

\subsection{Mixed Analysis}
The mixed analysis has almost become a necessity for all competitive teams in the CGC competition, which indicate that the mixed analysis is a powerful way in software security. There are many automatic methods proposed by researchers. DART\cite{godefroid2005dart},
CUTE\cite{sen2005cute},
EXE\cite{cadar2008exe},
KLEE\cite{cadar2008klee}
and SAGE\cite{godefroid2008automated}
are the typical representative of this area. The state-of-the-art technique, called concolic execution, is to have the concrete execution drive the symbolic execution. Concolic analysis maintains a concrete state and a symbolic state.
It executes a program starting with some given or random inputs that act on conditional statements to gather symbolic constraints along the execution. Then it uses a constraint solver to infer variants of the precious inputs in order to steer the next execution of the program towards an alternative execution path\cite{cadar2013symbolic}.

A key advantage of concolic analysis tools is that the use of concrete values can alleviate the imprecision caused by the interaction with external code or constraint solving timeouts. However they are still facing a serious challenge, that is, path explosion.
There are two ways to solve the problem: The first one is to combining online and offline (concolic) execution to obtain the best of both worlds. Mayhem\cite{cha2012unleashing} is a hybrid symbolic execution system that uses this way. It can context-switch quickly between states during the online exploration, while extraneous states are swapped to disk and explored in a new exploration\cite{avgerinos2014exploiting}.
The second one is to using concolic analysis to assist Fuzzing analysis. For example, Istvan Haller et al.\cite{haller2013dowser} develop Dowser, which proposes a new search algorithm with aiming to maximize pointer coverage to alleviate path explosion problem. And the typical and popular tool in this way is Driller proposed by Nick Stephens et al.\cite{stephens2016driller} that combined AFL and concolic analysis to solve the path explosion problem.
The above two ways attempt to go around path explosion by reducing the burden on the constraint solver. However it can not solve the problem fundamentally, and thus it would be a challenging topic for our future research.

\section{Automatic Vulnerability Exploitation}
Due to economic reasons, there are only a limited number of patches that can be actually deployed. Choosing which vulnerability to address becomes a critical challenge\cite{gianini2015game}. So the most important issue in software security is to find the security-critical bugs and fix them as soon as possible, which automatic vulnerability exploitation is in charge.

Automatic vulnerability exploits generation is an important and effective means to discover and verify bugs in a program. The existing methods can be classified into three categories roughly: Patch-Based Exploit Generation, Control Flow Hijacking Exploits Generation and Data-Oriented Exploits Generation\cite{WanYunpeng2017AEGSBoSE}.
This section will describe these three methods in detail.

\subsection{Patch-Based Exploits Generation}
A vulnerability often gets reproduced due to the frequent code reuse by programmers. Security patches are often not propagated to all code clones, but they could be leveraged to exploit unknown vulnerabilities\cite{li2014scalable}. Having early access to a patch may confer advantages to an attacker. A scheme of patch-based exploits generation is put forward by researchers in which the typical method is Automatic Patch-based Exploit Generation (APEG). APEG is proposed by David Brumley et al.\cite{brumley2008automatic}, and it is verified to be possible by successfully being used to generate exploits for 5 real-world vulnerabilities. However, from the practical point of view, the type of exploits generated by APEG is mainly denial of service, which can only cause the original program to crash, failing to result in a direct control flow hijacking.
Therefore, Control Flow Hijacking Exploit Generation is then put forward.

\subsection{Control Flow Hijacking Exploit Generation}
Techniques based on control flow hijacking have been proposed to overcome the APEG's defect of patch dependency and the inability to construct control flow hijacking.
Heelan\cite{heelan2009automatic} first implements the technique.
The classic method is Automatic Exploit Generation (AEG)\cite{thanassis2011aeg} proposed by Thanassis. AEG's core principle is mainly divided into four steps: Firstly, locate the location of the vulnerability. Secondly, get run-time information of the program such as the stack layout information. Thirdly, generate exploits based on the above information. Lastly, verify the exploits. AEG is the first truly automated solution for exploiting control flow oriented vulnerabilities. There are some limitations with AEG such as dependence on the source code, the exploit samples being limited by compilers, and the dynamic operating environments.

To address the above issues, Q\cite{schwartz2011q} later serves as an exploit hardening system proposed. Based on the research of AEG and Return-Oriented Programming (ROP)\cite{roemer2012return}, Q is designed to bypass the memory protection mechanisms like W$\bigoplus{X}$ and Address Space Layout Randomization (ASLR), etc.
PloyAEG\cite{wang2013automatic} is proposed that can automatically generate multiple exploits for a vulnerable program. Rex\cite{teamshellphish2017} tracks formulas for all registers and memory values to create a Proof Of Vulnerability (POV), and it is implemented in Mechanical Phish which finished third in the CGC Final Event in 2016.  All of them have improved AEG to some extent, making the automatic exploits generation technology more feasible. However, in the context of a broad deployment of defense solutions against control-flow hijacking attack, most of the attackers have turned their attention from control-oriented exploits towards data-oriented exploits.

\subsection{Data-Oriented Exploits Generation}
Data-oriented attacks change the path of data flow instead of changing the basic control flow of the program. Data-oriented attacks are not well known so far and there are few corresponding defense solutions. Therefore, data-oriented exploit generation has better applicability and flexibility.

Under such research background, the team of Liang put forward a new technique called data-flow stitching that is implemented by the tool FlowStitch\cite{hu2015automatic}. FlowStitch is the first scheme for automatic data-oriented exploits generation and verified to be feasible. However, it has some limitations, for example,
there must be at least one known memory error in the program as a precondition.
Later, they propose Data-Oriented Programming (DOP)\cite{hu2016data} that implements the method to search, extract and program for the attack code blocks and the instruction scheduling assignment code blocks in the actual applications. DOP is Turing-complete as it can execute arbitrary code and bypass the system defenses like DEP and ASLR. Above all, automatic vulnerability exploitation always requires deep semantic understanding that is difficult. So it is a hot topic and more researches are needed to enhance the machine analysis of deep semantic in the future.

\section{Automatic Vulnerability Patching}
Due to some realistic reasons, such as the rapid spread of worms\cite{Costa2005Vigilante}, researchers proposed many automatic patch technologies to ensure software's proper function in early years. These technologies are mainly used to prevent malware's hijack.
Until recently, automatic patch technology is gradually penetrating into almost all aspects of software.
In this section, we will introduce automatic vulnerability patching techniques that mainly focus on automatic patch technology in software security.

\subsection{The Automatic Patching Process}
The automatic patch technique commonly goes through several routine steps such as fault-locating, patch-generating, and function-verifying.

\subsubsection{fault-locating}
Fault-locating identifies the location of defects or failures. In the early stage, researchers located failures by monitoring programs, such as logging its running states, and then some researchers use fuzzing to find inputs that trigger failures. After that, static analysis methods, e.g., AST, are used. AST uses semantic analysis based methods to find software defects, and has good prospects. AST analyzes the structure of a program and provides information for patching.
Monitoring the execution of the program is still an effective choice for fault-locating when source code is not provided.

\subsubsection{Patch-generating}
Patch-generating is the most important step. There are two kinds of patches: vulnerability-tolerant and vulnerability-eliminating. For the vulnerability-tolerant patch, two patching strategies are widely used. The first is implementing an input filter to remove trigger-inputs. This strategy is used by VSEF\cite{Newsome2005Vulnerability} and TAP\cite{Sidiroglou2015Automatic}. The other is implementing a checkpoint mechanism. It is used to save states before executing vulnerable instructions and roll back to a safe state when an error occurs. For the vulnerability-eliminating patch, semantic-based patching method is proposed. This method called GenProg is first introduced by a team led by Goues C L. GenProg takes some patch templates, and searches a patch-space generated by semantic analysis for right patches.

\subsubsection{Function-verifying}
After patches are applied, the verification of the functionality of the patched software is required. Currently, the verification step is mainly conducted by software testing using a variety of test cases provided by the software designer or generated from software specification. After the occurrence of GenProg, this kind of verification technology becomes practical.

In terms of the way of vulnerability patching, this paper divides automatic patching methods into two categories, runtime state-based repair and detect-based repair. The following two subsections will describe these methods in detail.

\subsection{Runtime State-based Repair}
When a program is running, the following methods will repair the bugs or take some actions to bypass the vulnerabilities without discontinuing the program.

\subsubsection{Monitor}
To defense against the rapid spread of worms and other attacks, VSEF\cite{Newsome2005Vulnerability} combines automatic input filtering and binary-based full execution monitoring to achieve automatically patching of software.

DieHard\cite{Berger2006DieHard} aims at repairing heap overflow vulnerabilities. It uses randomization, replication and other methods to build an approximate infinite heap. Similar to DieHard, Exterminator\cite{Novark2007Exterminator} addresses buffer overflows and dangling pointers by exploiting randomization to pinpoint errors and merging patches to generate a new patch.

ClearView\cite{Perkins2009Automatically} is proposed by jeff et al.. It repairs bugs of binaries by observing the normal execution of registers and memory locations on x86 systems automatically. When errors are found, it
repairs errors through correcting the variables by comparing to normal runtime log. ClearView can alleviate memory write out of bounds vulnerabilities and invalid change of control flow vulnerabilities.

\subsubsection{Rollback}
Rx\cite{Qin2007Rx} is an automatic repair method proposed by F.Qin et al., which is inspired by the method of allergy treatment for removing allergens. Based on this idea, Rx changes the execution environment of the program back to health when occurring faults. Then Rx will restart the program, and try to rollback it to the checkpoint-state.

Different from the previous methods, ASSURE\cite{Sidiroglou2009ASSURE} presents the concept of Rescue Point (RP). RP is the location of a program that detects known errors and needs to restart. ASSURE can be regarded as an exception handling. It uses virtual execution to simulate the runtime environment and explores the reasons to repair the errors such as a memory write out of bounds and illegally control the transfer, etc. REASSURE\cite{Portokalidis2011REASSURE} is ASSURE-based with simplifying the deployment of ASSURE and improving its performance.

\subsection{Detect-based Repair}
Detect-based repair mainly uses fuzzing test to filter the input that can trigger the defect, or, uses AST to analyze and uses genetic programming to generate a patch to repair the program, or, repair the defect based on the constraint solving. Four ways of detect-based repair will be introduced in the following subsections.

\subsubsection{Input Filter}
Like other early developed automatic patch methods, Vigilante\cite{Costa2005Vigilante} is also designed to defense worms, and it generates input filters automatically by self-certifying alerts.
For buffer overflow, TAP\cite{Sidiroglou2015Automatic} designs two repair templates to recognize, generate and insert source-level patch automatically to filter input.

\subsubsection{Search-based Repair}
In recent years,
researchers devote genetic programming to automatic patching methods. GenProg\cite{Goues2012GenProg} is a success case, it uses genetic programming and extends it to repair program defects automatically without the need for formal specification, program annotations or special coding. It uses structural differencing algorithms and Delta debugging to further reduce the difference between the patched program and the original program, in order to achieve the automatic patching with minimal changes.

Pattern-based Automatic Program Repair (PAR)\cite{Kim2013Automatic}, uses repair patterns by learning from existing manual patches. Researchers have checked over 60,000 manual patches and found several common repair patterns for PAR to generate patches automatically. Although a large amount of time has been spent on manually creating the repair-templates, researchers declare that these templates have high reusability, and can be used on different occasions.

AE\cite{Forrest2013Leveraging} is based on GenProg, but it improves the performance of the healing process. Because AE formalizes repair cost in terms of test executions and uses many techniques such as syntax and data flow analysis to reduce patch search space. To improve the rate of success for patch generation, the order of tests should be adjusted. AE also sets the priority to patches which are the most likely to fail or to succeed.

Like AE,
SPR\cite{Long2015Staged} uses subsection repair and conditional synthesis to further reduce the search space and improve search efficiency. Propht\cite{Fan2016Automatic} learns from manual patches to create correct probabilistic models. Based on these probabilistic models, the test strategy will sort patches to identify the best patch for defects or vulnerabilities.

\subsubsection{Statistical Debugging Metrics}
Constraint solving has been applied to automatic patch technology in recent years. SemFix\cite{Nguyen2013SemFix} is a semantic-based automatic repair method that combines the symbolic execution, constraint solving, and program composition. In this method, programs that need to patch are formulated as constraints through a given set of tests. And then, SemFix synthesizes patches through semantic program analytics with dynamic symbolic execution.

DirectFix\cite{Mechtaev2015DirectFix}
and Angelix\cite{Mechtaev2016Angelix}
are based on SemFix. DirectFix differs from all existing repair methods, it combines the fault locating with the patch generating. DirectFix uses the MaxSMT solver to generate the minimized modification to the program.
Angelix is based on the angelic forest. Compared with GenProg and SPR, Angelix is scalable with the ability of high-quality and multi-line.

\subsubsection{Standard Semantics}
Because new errors will probably be added when repairing known errors, Gopinath et al.\cite{Gopinath2011Specification} propose a Specification-based code of conduct for automated patch technology, using regulation constraints to trim subsequent uncertainty and repair misstatements. They use the SAT-based Alloy tool-set to describe specification constraints and solve these constraints.

In terms of the above review, great progress has been made in automatic vulnerability patching. However, it cannot repair all types of bugs and it is challenging to fix zero-day vulnerabilities. As the most difficult part, automatic vulnerability patching requires better deep semantic understanding. Therefore the machine understanding of deep semantic can be regarded as the most challenging research topic, in which the machine learning will play an important role.

\section{Machine Learning in Software Security}
As mentioned in Sec.1, with the success of CGC in 2016, we are standing at a coming era in which the software vulnerability detection, exploitation and patching can be done by machines in an automated and scalable manner.  
Meanwhile, recent breakthroughs of Deep Learning (DL) in AlphaGo\cite{silver2016mastering} bring us new possibilities of utilizing DL-based or other machine learning-based technologies to improve the performance of automatic software vulnerability detection, exploitation and patching. In this section, we provide an extensive survey of the progress in this field, and discuss the open challenges.

\subsection{Vulnerability Detection and Exploitation}
Since vulnerability detection is very different from the typical problems for which machine learning is motivated, such as image classification, speech tagging, etc. The most important issue for utilizing machine learning to detect software vulnerabilities is to bridge this intention gap. To this end, some pioneer works have been published in recent years.

Some researchers try to import outside information that is needed during program execution to assist the vulnerability detecting process. For example, Perl et al.\cite{perl2015vccfinder} propose an approach that uses code repositories containing meta-data and code metrics to identify potentially vulnerable commits. They conduct a dataset consisting of 170,860 commits from 66 C/C++ GitHub projects, including 640 vulnerability-contributing commits mapped to the CVE IDs. Based on this dataset, heterogeneous features are extracted and jointly represented using a bag-of-words model. A Support Vector Machine (SVM) classifier is built on top of these features and trained using the dataset to detect the so-called vulnerability-contributing commits.

Some studies attempt to enhance deep semantic understanding. By extracting the semantic information
from the program to detect vulnerabilities,
Zhen Li et al. \cite{li2018vuldeepecker} implement a deep learning vulnerability detection system called Vulnerability Deep Pecker (VulDeePecker). Deep learning does not require the expert intervention to generate features, making the automation possible. However, applying deep learning to vulnerability detection still requires some guidance.
The authors use a code gadget that has multi-block code lines (not necessarily consecutive) with semantic dependencies to represent the software programs. A Bidirectional Long Short-Term Memory (Bi-LSTM) neural network is built up on top of these code gadgets and trained using the first vulnerabilities dataset designed for deep learning to detect. VulDeePecker detects four vulnerabilities that are not reported in the National Vulnerability Database (NVD) that proves its effectiveness.

For vulnerability exploitation, the semantic information is also crucial. Wei You et al. \cite{you2017semfuzz} find a new method called SemFuzz to automatically generate a proof-of-concept exploit by collecting the vulnerability-related texts published by CVE and Linux Git log.
Version information, vulnerability types, and vulnerability functions for the exploit are provided by CVE. Linux git logs provide a description of the vulnerability patch.
SemFuzz is a semantic-based fuzzing technology that uses natural-language processing to extract semantic information by analyzing vulnerability-related text rather than the code itself. Using such information, SemFuzz creates a call sequence to access vulnerable functions and use seed mutation as inputs for fuzzing until the vulnerability is triggered.
In order to test the validity of SemFuzz, the author collected more than 112 Linux kernel vulnerability reported by CVE over the past five years. 16\% of the vulnerabilities were detected and even zero-day vulnerabilities and undisclosed vulnerabilities are discovered.

\subsection{Vulnerability Patching}
The improvement of deep semantic information understanding is still at the core of vulnerability patching. Here are two main strategies: generate-and-validate approach and semantics-driven approach. Next, we briefly review the studies of machine learning applied to the vulnerability patching.

Ripon K. Saha et al.\cite{saha2017elixir} propose a generate-and-validate technique for object-oriented programs called ELIXIR. The key to this technique is to use method calls extensively, to build repair expressions, and to synthesize patches. The authors use machine learning to rank the concrete repairs, and the ensuing expansion of the repair space can be filtered. The features come from the repair semantic information like the code around the repair location.
The authors evaluate ELIXIR on two datasets: Defects4J and Bugs.jar. The results show that the number of fixed bug increased significantly under the condition of expanding the effective search and repair space.

Rahul Gupta et al.\cite{gupta2017deepfix} propose a semantics-driven and deep learning-based programming error repair technique called Deepfix. Deepfix fixes multiple programming errors by repeatedly calling trained neural networks. The authors have set up two types of datasets including correct and erroneous programs. The authors use the Gated Recurrent Unit (GRU) model to test 6971 erroneous programs to verify their method.

\subsection{Open Challenges}
Machine learning-based techniques can implement automation better, which indicates a new opportunity to harden the software security. However, we are still facing challenges when penetrating the machine learning into this field of software security. Here we list three typical challenges as follows.
\subsubsection{The deep semantic understanding}
According to the above review, we can learn that deep semantic information is crucial for software security. The better semantic understanding, the greater ability for the machine to achieve automatic vulnerability detection, exploitation and patching. So by utilizing machine learning to implement better deep semantic understanding is a hot research topic in the future.

\subsubsection{The binary-based machine learning}
Shin et al. \cite{shin2015recognizing} apply the method of recursive neural networks to binary analysis, but it is not directly applied for vulnerability detection. Most machine learning algorithms are designed with source code while binary-based machine learning is rare. Therefore, binary-oriented machine learning vulnerability detection still needs to be explored.

\subsubsection{The open source datasets}
The existing datasets are collected or constructed temporarily according to some specific purposes. It lacks open source datasets like ImageNet\cite{ImageNet2018} that has a wide coverage and has a large amount of semantic information such as comprehensive data type. More importantly, it can be used for vulnerability detection and patching based on machine learning. Thus the open source datasets are necessary.

\section{Conclusion}
With the development of the Internet, the security issues caused by software are spreading widely. The autonomous CRS begin to attract our attention. Many research scholars have put forward a variety of related techniques attempting to replace human labor. This paper provides a detailed introduction to recent research technologies in the areas of automatic vulnerability detection, exploitation and patching. Moreover, machine learning in software security is playing an important role. This paper hereby carried out the latest research works in the field of machine learning. It also summarizes some challenges of the machine learning in software security. And in the future, it is promising to conduct the research work from the perspective of machine learning on the application of automatic vulnerability detection, exploitation and patching.

\bibliographystyle{IEEEtran}
\bibliography{myconference}

\begin{thebibliography}{10}
\providecommand{\url}[1]{#1}
\csname url@samestyle\endcsname
\providecommand{\newblock}{\relax}
\providecommand{\bibinfo}[2]{#2}
\providecommand{\BIBentrySTDinterwordspacing}{\spaceskip=0pt\relax}
\providecommand{\BIBentryALTinterwordstretchfactor}{4}
\providecommand{\BIBentryALTinterwordspacing}{\spaceskip=\fontdimen2\font plus
\BIBentryALTinterwordstretchfactor\fontdimen3\font minus
  \fontdimen4\font\relax}
\providecommand{\BIBforeignlanguage}[2]{{%
\expandafter\ifx\csname l@#1\endcsname\relax
\typeout{** WARNING: IEEEtran.bst: No hyphenation pattern has been}%
\typeout{** loaded for the language `#1'. Using the pattern for}%
\typeout{** the default language instead.}%
\else
\language=\csname l@#1\endcsname
\fi
#2}}
\providecommand{\BIBdecl}{\relax}
\BIBdecl

\bibitem{CVE2018}
\BIBentryALTinterwordspacing
T.~M. Corporation. (2018, Jan.) Common vulnerabilities and exposures (cve®).
  [Online]. Available: \url{https://cve.mitre.org/}
\BIBentrySTDinterwordspacing

\bibitem{avgerinos2014exploiting}
A.~Avgerinos, ``Exploiting trade-offs in symbolic execution for identifying
  security bugs,'' 2014.

\bibitem{silver2016mastering}
D.~Silver, A.~Huang, C.~J. Maddison, A.~Guez, L.~Sifre, G.~Van Den~Driessche,
  J.~Schrittwieser, I.~Antonoglou, V.~Panneershelvam, M.~Lanctot \emph{et~al.},
  ``Mastering the game of go with deep neural networks and tree search,''
  \emph{nature}, vol. 529, no. 7587, pp. 484--489, 2016.

\bibitem{brooks2017survey}
T.~N. Brooks, ``Survey of automated vulnerability detection and exploit
  generation techniques in cyber reasoning systems,'' \emph{arXiv preprint
  arXiv:1702.06162}, 2017.

\bibitem{shoshitaishvili2017building}
Y.~Shoshitaishvili, ``Building a base for cyber-autonomy,'' Ph.D. dissertation,
  University of California, Santa Barbara, 2017.

\bibitem{song2008bitblaze}
D.~Song, D.~Brumley, H.~Yin, J.~Caballero, I.~Jager, M.~Kang, Z.~Liang,
  J.~Newsome, P.~Poosankam, and P.~Saxena, ``Bitblaze: A new approach to
  computer security via binary analysis,'' \emph{Information systems security},
  pp. 1--25, 2008.

\bibitem{yamaguchi2014modeling}
F.~Yamaguchi, N.~Golde, D.~Arp, and K.~Rieck, ``Modeling and discovering
  vulnerabilities with code property graphs,'' in \emph{Security and Privacy
  (SP), 2014 IEEE Symposium on}.\hskip 1em plus 0.5em minus 0.4em\relax IEEE,
  2014, pp. 590--604.

\bibitem{shoshitaishvili2016sok}
Y.~Shoshitaishvili, R.~Wang, C.~Salls, N.~Stephens, M.~Polino, A.~Dutcher,
  J.~Grosen, S.~Feng, C.~Hauser, C.~Kruegel \emph{et~al.}, ``Sok:(state of) the
  art of war: Offensive techniques in binary analysis,'' in \emph{Security and
  Privacy (SP), 2016 IEEE Symposium on}.\hskip 1em plus 0.5em minus 0.4em\relax
  IEEE, 2016, pp. 138--157.

\bibitem{cousot1977abstract}
P.~Cousot and R.~Cousot, ``Abstract interpretation: a unified lattice model for
  static analysis of programs by construction or approximation of fixpoints,''
  in \emph{Proceedings of the 4th ACM SIGACT-SIGPLAN symposium on Principles of
  programming languages}.\hskip 1em plus 0.5em minus 0.4em\relax ACM, 1977, pp.
  238--252.

\bibitem{balakrishnan2008wysinwyx}
G.~Balakrishnan, T.~Reps, D.~Melski, and T.~Teitelbaum, ``Wysinwyx: What you
  see is not what you execute,'' \emph{Verified software: theories, tools,
  experiments}, pp. 202--213, 2008.

\bibitem{cheng2011loongchecker}
S.~Cheng, J.~Yang, J.~Wang, J.~Wang, and F.~Jiang, ``Loongchecker: Practical
  summary-based semi-simulation to detect vulnerability in binary code,'' in
  \emph{Trust, Security and Privacy in Computing and Communications (TrustCom),
  2011 IEEE 10th International Conference on}.\hskip 1em plus 0.5em minus
  0.4em\relax IEEE, 2011, pp. 150--159.

\bibitem{feist2014statically}
J.~Feist, L.~Mounier, and M.-L. Potet, ``Statically detecting use after free on
  binary code,'' \emph{Journal of Computer Virology and Hacking Techniques},
  vol.~10, no.~3, pp. 211--217, 2014.

\bibitem{li2014scalable}
H.~Li, H.~Kwon, J.~Kwon, and H.~Lee, ``A scalable approach for vulnerability
  discovery based on security patches,'' in \emph{International Conference on
  Applications and Techniques in Information Security}.\hskip 1em plus 0.5em
  minus 0.4em\relax Springer, 2014, pp. 109--122.

\bibitem{jianming2014pvdf}
F.~Jianming, C.~Jing, P.~Guojun \emph{et~al.}, ``Pvdf: an automatic patch-based
  vulnerability description and fuzzing method,'' 2014.

\bibitem{xu2017spain}
Z.~Xu, B.~Chen, M.~Chandramohan, Y.~Liu, and F.~Song, ``Spain: security patch
  analysis for binaries towards understanding the pain and pills,'' in
  \emph{Proceedings of the 39th International Conference on Software
  Engineering}.\hskip 1em plus 0.5em minus 0.4em\relax IEEE Press, 2017, pp.
  462--472.

\bibitem{sutton2007fuzzing}
M.~Sutton, A.~Greene, and P.~Amini, \emph{Fuzzing: brute force vulnerability
  discovery}.\hskip 1em plus 0.5em minus 0.4em\relax Pearson Education, 2007.

\bibitem{godefroid2008automated}
P.~Godefroid, M.~Y. Levin, D.~A. Molnar \emph{et~al.}, ``Automated whitebox
  fuzz testing.'' in \emph{NDSS}, vol.~8, 2008, pp. 151--166.

\bibitem{bohme2017coverage}
M.~B{\"o}hme, V.-T. Pham, and A.~Roychoudhury, ``Coverage-based greybox fuzzing
  as markov chain,'' \emph{IEEE Transactions on Software Engineering}, 2017.

\bibitem{bohme2017directed}
M.~B{\"o}hme, V.-T. Pham, M.-D. Nguyen, and A.~Roychoudhury, ``Directed greybox
  fuzzing,'' in \emph{Proceedings of the 24th ACM Conference on Computer and
  Communications Security, ser. CCS}, 2017, pp. 1--16.

\bibitem{denning1976lattice}
D.~E. Denning, ``A lattice model of secure information flow,''
  \emph{Communications of the ACM}, vol.~19, no.~5, pp. 236--243, 1976.

\bibitem{newsome2005dynamic}
J.~Newsome and D.~Song, ``Dynamic taint analysis for automatic detection,
  analysis, and signature generation of exploits on commodity software,'' 2005.

\bibitem{schwartz2010all}
E.~J. Schwartz, T.~Avgerinos, and D.~Brumley, ``All you ever wanted to know
  about dynamic taint analysis and forward symbolic execution (but might have
  been afraid to ask),'' in \emph{Security and privacy (SP), 2010 IEEE
  symposium on}.\hskip 1em plus 0.5em minus 0.4em\relax IEEE, 2010, pp.
  317--331.

\bibitem{kang2011dta++}
M.~G. Kang, S.~McCamant, P.~Poosankam, and D.~Song, ``Dta++: dynamic taint
  analysis with targeted control-flow propagation.'' in \emph{NDSS}, 2011.

\bibitem{henderson2017decaf}
A.~Henderson, L.~K. Yan, X.~Hu, A.~Prakash, H.~Yin, and S.~McCamant, ``Decaf: A
  platform-neutral whole-system dynamic binary analysis platform,'' \emph{IEEE
  Transactions on Software Engineering}, vol.~43, no.~2, pp. 164--184, 2017.

\bibitem{MaJinXin2017}
J.~X. Ma, Z.~J. Li, T.~Zhang, D.~Shen, and Z.~K. Zhang, ``Taint analysis method
  based on offline indices of instruction trace,'' \emph{Ruan Jian Xue
  Bao/Journal of Software}, vol.~28, no.~9.

\bibitem{godefroid2005dart}
P.~Godefroid, N.~Klarlund, and K.~Sen, ``Dart: directed automated random
  testing,'' in \emph{ACM Sigplan Notices}, vol.~40, no.~6.\hskip 1em plus
  0.5em minus 0.4em\relax ACM, 2005, pp. 213--223.

\bibitem{sen2005cute}
K.~Sen, D.~Marinov, and G.~Agha, ``Cute: a concolic unit testing engine for
  c,'' in \emph{ACM SIGSOFT Software Engineering Notes}, vol.~30, no.~5.\hskip
  1em plus 0.5em minus 0.4em\relax ACM, 2005, pp. 263--272.

\bibitem{cadar2008exe}
C.~Cadar, V.~Ganesh, P.~M. Pawlowski, D.~L. Dill, and D.~R. Engler, ``Exe:
  automatically generating inputs of death,'' \emph{ACM Transactions on
  Information and System Security (TISSEC)}, vol.~12, no.~2, p.~10, 2008.

\bibitem{cadar2008klee}
C.~Cadar, D.~Dunbar, D.~R. Engler \emph{et~al.}, ``Klee: Unassisted and
  automatic generation of high-coverage tests for complex systems programs.''
  in \emph{OSDI}, vol.~8, 2008, pp. 209--224.

\bibitem{cadar2013symbolic}
C.~Cadar and K.~Sen, ``Symbolic execution for software testing: three decades
  later,'' \emph{Communications of the ACM}, vol.~56, no.~2, pp. 82--90, 2013.

\bibitem{cha2012unleashing}
S.~K. Cha, T.~Avgerinos, A.~Rebert, and D.~Brumley, ``Unleashing mayhem on
  binary code,'' in \emph{Security and Privacy (SP), 2012 IEEE Symposium
  on}.\hskip 1em plus 0.5em minus 0.4em\relax IEEE, 2012, pp. 380--394.

\bibitem{haller2013dowser}
I.~Haller, A.~Slowinska, M.~Neugschwandtner, and H.~Bos, ``Dowser: a guided
  fuzzer to find buffer overflow vulnerabilities,'' in \emph{Proceedings of the
  22nd USENIX Security Symposium}, 2013, pp. 49--64.

\bibitem{stephens2016driller}
N.~Stephens, J.~Grosen, C.~Salls, A.~Dutcher, R.~Wang, J.~Corbetta,
  Y.~Shoshitaishvili, C.~Kruegel, and G.~Vigna, ``Driller: Augmenting fuzzing
  through selective symbolic execution.'' in \emph{NDSS}, vol.~16, 2016, pp.
  1--16.

\bibitem{gianini2015game}
G.~Gianini, M.~Cremonini, A.~Rainini, G.~L. Cota, and L.~G. Fossi, ``A game
  theoretic approach to vulnerability patching,'' in \emph{Information and
  Communication Technology Research (ICTRC), 2015 International Conference
  on}.\hskip 1em plus 0.5em minus 0.4em\relax IEEE, 2015, pp. 88--91.

\bibitem{WanYunpeng2017AEGSBoSE}
Y.~P. Wan, Y.~Deng, D.~H. Shi, L.~Cheng, and Y.~Zhang, ``Automatic exploit
  generation system based on symbolic execution,'' \emph{Computer Systems {\&}
  Applications}, vol.~26, no.~10, pp. 44--52, 2017.

\bibitem{brumley2008automatic}
D.~Brumley, P.~Poosankam, D.~Song, and J.~Zheng, ``Automatic patch-based
  exploit generation is possible: Techniques and implications,'' in
  \emph{Security and Privacy, 2008. SP 2008. IEEE Symposium on}.\hskip 1em plus
  0.5em minus 0.4em\relax IEEE, 2008, pp. 143--157.

\bibitem{heelan2009automatic}
S.~Heelan, ``Automatic generation of control flow hijacking exploits for
  software vulnerabilities,'' Ph.D. dissertation, University of Oxford, 2009.

\bibitem{thanassis2011aeg}
H.~A. Thanassis, C.~S. Kil, and B.~David, ``Aeg: Automatic exploit
  generation,'' in \emph{ser. Network and Distributed System Security
  Symposium}, 2011.

\bibitem{schwartz2011q}
E.~J. Schwartz, T.~Avgerinos, and D.~Brumley, ``Q: Exploit hardening made
  easy.'' in \emph{USENIX Security Symposium}, 2011, pp. 25--41.

\bibitem{roemer2012return}
R.~Roemer, E.~Buchanan, H.~Shacham, and S.~Savage, ``Return-oriented
  programming: Systems, languages, and applications,'' \emph{ACM Transactions
  on Information and System Security (TISSEC)}, vol.~15, no.~1, p.~2, 2012.

\bibitem{wang2013automatic}
M.~Wang, P.~Su, Q.~Li, L.~Ying, Y.~Yang, and D.~Feng, ``Automatic polymorphic
  exploit generation for software vulnerabilities,'' in \emph{International
  Conference on Security and Privacy in Communication Systems}.\hskip 1em plus
  0.5em minus 0.4em\relax Springer, 2013, pp. 216--233.

\bibitem{teamshellphish2017}
\BIBentryALTinterwordspacing
T.~Shellphish. (2017, Jan.) Cyber grand shellphish. [Online]. Available:
  \url{http://phrack.org/papers/cyber{\_}grand{\_}shellphish.html}
\BIBentrySTDinterwordspacing

\bibitem{hu2015automatic}
H.~Hu, Z.~L. Chua, S.~Adrian, P.~Saxena, and Z.~Liang, ``Automatic generation
  of data-oriented exploits.'' in \emph{USENIX Security Symposium}, 2015, pp.
  177--192.

\bibitem{hu2016data}
H.~Hu, S.~Shinde, S.~Adrian, Z.~L. Chua, P.~Saxena, and Z.~Liang,
  ``Data-oriented programming: On the expressiveness of non-control data
  attacks,'' in \emph{Security and Privacy (SP), 2016 IEEE Symposium on}.\hskip
  1em plus 0.5em minus 0.4em\relax IEEE, 2016, pp. 969--986.

\bibitem{Costa2005Vigilante}
M.~Costa, J.~Crowcroft, M.~Castro, A.~Rowstron, L.~Zhou, L.~Zhang, and
  P.~Barham, ``Vigilante: end-to-end containment of internet worms,'' in
  \emph{Twentieth ACM Symposium on Operating Systems Principles}, 2005, pp.
  133--147.

\bibitem{Newsome2005Vulnerability}
J.~Newsome, D.~Brumley, and D.~X. Song, ``Vulnerability-specific execution
  filtering for exploit prevention on commodity software,'' in \emph{Network
  and Distributed System Security Symposium, NDSS 2006, San Diego, California,
  Usa}, 2005.

\bibitem{Sidiroglou2015Automatic}
S.~Sidiroglou-Douskos, E.~Lahtinen, and M.~Rinard, ``Automatic discovery and
  patching of buffer and integer overflow errors,'' 2015.

\bibitem{Berger2006DieHard}
E.~D. Berger and B.~G. Zorn, ``Diehard: Probabilistic memory safety for unsafe
  languages,'' \emph{Acm Sigplan Notices}, vol.~41, no.~6, pp. 158--168, 2006.

\bibitem{Novark2007Exterminator}
G.~Novark, E.~D. Berger, and B.~G. Zorn, ``Exterminator: automatically
  correcting memory errors with high probability,'' in \emph{ACM Sigplan
  Conference on Programming Language Design and Implementation}, 2007, pp.
  1--11.

\bibitem{Perkins2009Automatically}
J.~H. Perkins, S.~Kim, S.~Larsen, S.~Amarasinghe, J.~Bachrach, M.~Carbin,
  F.~Sherwood, F.~Sherwood, G.~Sullivan, and G.~Sullivan, ``Automatically
  patching errors in deployed software,'' in \emph{ACM Sigops Symposium on
  Operating Systems Principles}, 2009, pp. 87--102.

\bibitem{Qin2007Rx}
F.~Qin, J.~Tucek, J.~Sundaresan, and J.~Sundaresan, ``Rx: Treating bugs as
  allergies—a safe method to survive software failures,'' \emph{ACM
  Transactions on Computer Systems (TOCS)}, vol.~25, no.~3, p.~7, 2007.

\bibitem{Sidiroglou2009ASSURE}
S.~Sidiroglou, O.~Laadan, C.~Perez, N.~Viennot, J.~Nieh, and A.~D. Keromytis,
  ``Assure: automatic software self-healing using rescue points,'' \emph{Acm
  Sigarch Computer Architecture News}, vol.~37, no.~1, pp. 37--48, 2009.

\bibitem{Portokalidis2011REASSURE}
G.~Portokalidis and A.~D. Keromytis, ``Reassure: A self-contained mechanism for
  healing software using rescue points,'' in \emph{International Conference on
  Advances in Information and Computer Security}, 2011, pp. 16--32.

\bibitem{Goues2012GenProg}
C.~L. Goues, T.~V. Nguyen, S.~Forrest, and W.~Weimer, ``Genprog: A generic
  method for automatic software repair,'' \emph{IEEE Transactions on Software
  Engineering}, vol.~38, no.~1, pp. 54--72, 2012.

\bibitem{Kim2013Automatic}
D.~Kim, J.~Nam, J.~Song, and S.~Kim, ``Automatic patch generation learned from
  human-written patches,'' in \emph{International Conference on Software
  Engineering}, 2013, pp. 802--811.

\bibitem{Forrest2013Leveraging}
S.~Forrest, S.~Forrest, and S.~Forrest, ``Leveraging program equivalence for
  adaptive program repair: models and first results,'' in \emph{Ieee/acm
  International Conference on Automated Software Engineering}, 2013, pp.
  356--366.

\bibitem{Long2015Staged}
F.~Long and M.~Rinard, ``Staged program repair with condition synthesis,'' in
  \emph{Joint Meeting on Foundations of Software Engineering}, 2015, pp.
  166--178.

\bibitem{Fan2016Automatic}
L.~Fan and M.~Rinard, ``Automatic patch generation by learning correct code,''
  2016, pp. 298--312.

\bibitem{Nguyen2013SemFix}
H.~D.~T. Nguyen, D.~Qi, A.~Roychoudhury, and S.~Chandra, ``Semfix: Program
  repair via semantic analysis,'' in \emph{International Conference on Software
  Engineering}, 2013, pp. 772--781.

\bibitem{Mechtaev2015DirectFix}
S.~Mechtaev, J.~Yi, and A.~Roychoudhury, ``Directfix: looking for simple
  program repairs,'' in \emph{Ieee/acm IEEE International Conference on
  Software Engineering}, 2015, pp. 448--458.

\bibitem{Mechtaev2016Angelix}
------, ``Angelix: Scalable multiline program patch synthesis via symbolic
  analysis,'' in \emph{Ieee/acm International Conference on Software
  Engineering}, 2016, pp. 691--701.

\bibitem{Gopinath2011Specification}
D.~Gopinath, M.~Z. Malik, and S.~Khurshid, \emph{Specification-Based Program
  Repair Using SAT}.\hskip 1em plus 0.5em minus 0.4em\relax Springer Berlin
  Heidelberg, 2011.

\bibitem{perl2015vccfinder}
H.~Perl, S.~Dechand, M.~Smith, D.~Arp, F.~Yamaguchi, K.~Rieck, S.~Fahl, and
  Y.~Acar, ``Vccfinder: Finding potential vulnerabilities in open-source
  projects to assist code audits,'' in \emph{Proceedings of the 22nd ACM SIGSAC
  Conference on Computer and Communications Security}.\hskip 1em plus 0.5em
  minus 0.4em\relax ACM, 2015, pp. 426--437.

\bibitem{li2018vuldeepecker}
Z.~Li, D.~Zou, S.~Xu, X.~Ou, H.~Jin, S.~Wang, Z.~Deng, and Y.~Zhong,
  ``Vuldeepecker: A deep learning-based system for vulnerability detection,''
  in \emph{25th Annual Network and Distributed System Security Symposium (NDSS
  2018), San Diego, California, USA, February 18-21, 2018 (EI/CCF-A)}, 2018.

\bibitem{you2017semfuzz}
W.~You, P.~Zong, K.~Chen, X.~Wang, X.~Liao, P.~Bian, and B.~Liang, ``Semfuzz:
  Semantics-based automatic generation of proof-of-concept exploits,'' in
  \emph{Proceedings of the 2017 ACM SIGSAC Conference on Computer and
  Communications Security}.\hskip 1em plus 0.5em minus 0.4em\relax ACM, 2017,
  pp. 2139--2154.

\bibitem{saha2017elixir}
R.~K. Saha, Y.~Lyu, H.~Yoshida, and M.~R. Prasad, ``Elixir: effective object
  oriented program repair,'' in \emph{Proceedings of the 32nd IEEE/ACM
  International Conference on Automated Software Engineering}.\hskip 1em plus
  0.5em minus 0.4em\relax IEEE Press, 2017, pp. 648--659.

\bibitem{gupta2017deepfix}
R.~Gupta, S.~Pal, A.~Kanade, and S.~Shevade, ``Deepfix: Fixing common c
  language errors by deep learning.'' in \emph{AAAI}, 2017, pp. 1345--1351.

\bibitem{shin2015recognizing}
E.~C.~R. Shin, D.~Song, and R.~Moazzezi, ``Recognizing functions in binaries
  with neural networks.'' in \emph{USENIX Security Symposium}, 2015, pp.
  611--626.

\bibitem{ImageNet2018}
\BIBentryALTinterwordspacing
P.~U. 2016 Stanford Vision~Lab, Stanford~University. (2018, May) Imagenet.
  [Online]. Available: \url{http://www.image-net.org/}
\BIBentrySTDinterwordspacing

\end{thebibliography}

\end{CJK*}
\end{document}